\documentclass[10pt,conference,letterpaper]{IEEEtran}
\usepackage{times}
\usepackage{epsfig}
\usepackage{psfrag}
\usepackage{amsmath}
\usepackage{amssymb}
\usepackage{amsfonts}
\usepackage{color}
\usepackage{times}

\newcommand{\argmax}{\mathop{\mathrm{argmax}}}
\newcommand{\argmin}{\mathop{\mathrm{argmin}}}
\def\PSNR{\mathrm{ PSNR}}

\def\punit{\, \mathrm}

\title{Multiple Selection Extrapolation for Improved Spatial Error Concealment\vspace*{-0.1cm}}
\author{
{J\"urgen~Seiler and Andr\'e~Kaup}%
\vspace{1.6mm}\\
\fontsize{10}{10}\selectfont\itshape Chair of Multimedia Communications and Signal Processing, \\University of Erlangen-Nuremberg, Cauerstr. 7, 91058 Erlangen, Germany\\ {\fontsize{9}{9}\selectfont\ttfamily\upshape \{seiler, kaup\}@LNT.de}\vspace*{-0.3cm}
}

\begin{document}
\maketitle

\begin{abstract} 
This contribution introduces a novel signal extrapolation algorithm and its application to image error concealment. The signal extrapolation is carried out by iteratively generating a model of the signal suffering from distortion. Thereby, the model results from a weighted superposition of two-dimensional basis functions whereas in every iteration step a set of these is selected and the approximation residual is projected onto the subspace they span. The algorithm is an improvement to the Frequency Selective Extrapolation that has proven to be an effective method for concealing lost or distorted image regions. Compared to this algorithm, the novel algorithm is able to reduce the processing time by a factor larger than three, by still preserving the very high extrapolation quality. 
\end{abstract}


\section{Introduction}
\label{sec:introduction}

Signal extrapolation is an important signal processing task and is often used in image and video processing. Thereby a signal gets extended from a limited number of known samples into areas where no knowledge of the signal is existent. A problem, signal extrapolation is applied to quite often, is the concealment of losses during image and video communication. In the case that transmission errors occur while transmitting the bit stream of a coded image or a video sequence, the received signal cannot be decoded correctly and several areas of the image or the sequence get distorted or lost. In order to conceal these errors or losses, the signal can be extra\-polated from neighboring, correctly received areas into the area where the distortion occurred. Besides error concealment signal extrapolation can also be used e.\ g. for prediction in video coding. Thereby, the block actually being coded is predicted from already transmitted macroblocks, so only the prediction error has to be transmitted. Thus, the amount of data to be transported directly depends on the prediction and therewith the extrapolation quality. Subsequently, a novel signal extrapolation algorithm will be presented at the example of spatial error concealment.

For the problem of spatial error concealment a wide range of extrapolation algorithms exist. Without claiming to be a complete list, some widely known algorithms should be mentioned. There are e.\ g. the maximally smooth image recovery from Wang et al. \cite{Wang1993}, the DCT-based interpolation proposed by Alkachouh and Bellanger \cite{Alkachouh2000}, the utilization of projection onto convex sets introduced by Sun and Kwok \cite{Sun1995}, the sequential error concealment from Li and Orchard \cite{Li2002} or the error concealment by directional interpolation proposed by Zhao et al. \cite{Zhao2008}. Another, very effective error concealment algorithm is the Frequency Selective Extrapolation (FSE) which was originally presented in \cite{Meisinger2004b}, and its enhancement presented in \cite{Seiler2008} respectively. This algorithm iteratively generates a model of the signal, emanating from a weighted superposition of two-dimensional basis functions. The model is generated in order to approximate the signal in the undistorted areas. Since the model is defined over the lost areas as well, the signal gets extrapolated into these areas. This algorithm is well suited for extrapolating smooth as well as noise like signals and edges and it outperforms most other error concealment algorithms. It has however a shortcoming: the algorithm is computationally very expensive. Since in every iteration step only one basis function is added to the model, many iterations are necessary for generating the model. 

To cope with this, we want to propose a novel extrapolation algorithm, the Multiple Selection Extrapolation (MuSE). MuSE is an enhancement to FSE and is able to generate the model more efficiently. This is done by selecting several basis functions to be added to the model in one iteration step and by using an advanced update of the signal model. In doing so, the computational cost for the individual iteration steps is slightly increased whereas the overall number of iterations and therewith the overall computational cost is considerably reduced. Besides the reduced computational cost, the novel algorithm can be adjusted more easily to different problems by being able to satisfy runtime or quality constraints.

In the next section, the FSE is briefly surveyed in order to introduce the extrapolation scenario and the original algorithm. Afterwards, the novel MuSE is presented in detail and the advantages over FSE are carried out.


\section{Frequency Selective Extrapolation}
\label{sec:fsextrapolation}

\begin{figure}
	\psfrag{m}[c][c][0.8]{$m$}
	\psfrag{n}[c][c][0.8]{$n$}
	\psfrag{M}[c][c][0.8]{$M$}
	\psfrag{N}[c][c][0.8]{$N$}
	\psfrag{L}[l][l][0.8]{$\mathcal{L} = \mathcal{A} \cup \mathcal{B}$}
	\psfrag{A}[l][l][0.8]{$\mathcal{A}$}
	\psfrag{B}[l][l][0.8]{$\mathcal{B}$}
	\centering
	\includegraphics[width=0.16\textwidth]{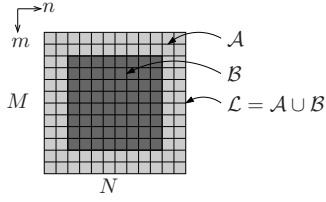}
	\caption{Loss scenario used for signal extrapolation. The data area $\mathcal{L}$ used for two-dimensional extrapolation consists of the loss area $\mathcal{B}$ to be estimated and the known surrounding support area $\mathcal{A}$.\vspace{-2.5mm}}
	\label{fig:loss_scenario}
\end{figure}

Frequency Selective Extrapolation (FSE) is an iterative signal extrapolation algorithm. Without loss of generality, here an extrapolation scenario as shown in Fig.\ \ref{fig:loss_scenario} is assumed. We regard the data area $\mathcal{L}$ of size $M\times N$ samples with spatial coordinates $m$ and $n$. Area $\mathcal{L}$ consists of two sub-areas, the loss area $\mathcal{B}$ and the support area $\mathcal{A}$. We further regard the signal $f\left[m,n\right]$. It is defined over the complete area $\mathcal{L}$, but its magnitude is only accessible over the support area $\mathcal{A}$ as the signal information is lost in area $\mathcal{B}$. FSE now aims at extrapolating the signal from area $\mathcal{A}$ into area $\mathcal{B}$. For this, FSE generates the parametric model $g\left[m,n\right]$ that is defined over the complete area $\mathcal{L}$ and therewith continues the signal into area $\mathcal{B}$.  The parametric model 
\begin{equation}
 	g\left[m,n\right] = \sum_{\forall k\in \mathfrak{K}} c_k \varphi_k\left[m,n\right]
\end{equation}
emerges from a weighted superposition of mutually orthogonal two-dimensional basis functions depicted by $\varphi_k\left[m,n\right]$. The set $\mathfrak{K}$ covers the indices of all basis functions used for the model generation. The weighting factors are depicted by $c_k$ and are called expansion coefficients. 

The actual model generation is conducted in such a way that the generated model approximates $f\left[m,n\right]$ in the support area $\mathcal{A}$. Since the basis functions are defined over complete $\mathcal{L}$, the signal gets extrapolated into area $\mathcal{B}$. As mentioned above, FSE is an iterative algorithm, whereas in every iteration step one basis function is selected to be added to the model generated so far. Thus, the model in the $\nu$-th iteration step is
\begin{equation}
 	\label{eq:fse_model_update}
	g^{\left(\nu\right)}\left[m,n\right] = g^{\left(\nu-1\right)}\left[m,n\right] + \hat{c}^{\left(\nu\right)}_u \cdot \varphi_u\left[m,n\right] 
\end{equation}
with $u$ being the index of the selected basis function and $\hat{c}^{\left(\nu\right)}_u$ being the update of the corresponding expansion coefficient in this iteration step. The initial model $g^{\left(0\right)}\left[m,n\right]$ is set to $0$. In the $\nu$-th iteration step the approximation error $r^{\left(\nu\right)}\left[m,n\right]$ between $f\left[m,n\right]$ and $g^{\left(\nu\right)}\left[m,n\right]$ can be calculated according to
\begin{equation}
 	\hspace{-2mm}r^{\left(\nu\right)}\hspace{-1mm}\left[m,n\right] \hspace{-1mm}=\hspace{-1mm} \left\{\hspace{-1.75mm} \begin{array}{ll}  r^{\left(\nu-1\right)}\hspace{-1mm}\left[m,n\right] \hspace{-0.5mm}-\hspace{-0.5mm} \hat{c}^{\left(\nu\right)}_u \varphi_u\hspace{-1mm}\left[m,n\right] \hspace{-2.5mm} & , \forall \left(m,n\right) \in \mathcal{A} \\ 0 & , \forall \left(m,n\right) \in \mathcal{B} \end{array} \right.
\end{equation}
since the error can only be evaluated over the support area. 

For determining the basis function to be added to the model in a certain iteration step, a weighted projection of $r^{\left(\nu-1\right)}\left[m,n\right]$ onto all basis functions is performed. This yields the projection coefficients 
\begin{equation}
\label{eq:projection}
 	p_k^{\left(\nu\right)} = \frac{\displaystyle \sum_{\left(m,n\right) \in \mathcal{L}}r^{\left(\nu-1\right)}\left[m,n\right] \cdot \varphi_k\left[m,n\right] \cdot w\left[m,n\right]} {\displaystyle \sum_{\left(m,n\right) \in \mathcal{L}}\varphi^2_k\left[m,n\right] \cdot w\left[m,n\right]}  
\end{equation}
resulting from the weighted projection of $r^{\left(\nu-1\right)}\left[m,n\right]$ onto the $k$-th basis function. The weighting function 
\begin{equation}
 	w\left[m,n\right] = \left\{ \begin{array}{ll} \rho\left[m,n\right] & , \forall \left(m,n\right) \in \mathcal{A} \\ 0& , \forall \left(m,n\right) \in \mathcal{B} \end{array} \right.
\end{equation}
is used to mask area $\mathcal{B}$, since there the weighted scalar product between $r^{\left(\nu-1\right)}\left[m,n\right]$ and $\varphi_k\left[m,n\right]$ cannot be evaluated. In addition to this, the arbitrary function $\rho\left[m,n\right]$ controls the influence samples have on the model generation, depending on their position. So samples close to $\mathcal{B}$ can e.\ g.\ get a higher impact on the model generation than the ones farther away. 

After all $p_k^{\left(\nu\right)}$ have been calculated, the basis function to be added to the model in this iteration step is determined. There, the one is chosen that minimizes the distance between $r^{\left(\nu-1\right)}\left[m,n\right]$ and the projection onto the basis function. This is also the basis function that maximizes the decrement of the weighted approximation error energy. The index $u$ can be determined according to
\begin{eqnarray}
 	\nonumber u \hspace{-2mm} &=& \hspace{-2mm} \argmin_k\sum_{\left(m,n\right)\in \mathcal{L}}\hspace{-1mm} \left(r^{\left(\nu-1\right)}\left[m,n\right]\hspace{-0.5mm} -\hspace{-0.5mm} p_k^{\left(\nu\right)} \varphi_k\left[m,n\right]\right)^2 \hspace{-1mm}w\left[m,n\right] \\
	&=& \hspace{-2mm}\argmax_k \ \ p_k^{\left(\nu\right)^2} \cdot \sum_{\left(m,n\right)\in \mathcal{L}} \varphi_k^2\left[m,n\right] w\left[m,n\right] .
\end{eqnarray}

After having determined the basis function to be added, the corresponding expansion coefficient $\hat{c}_u^{\left(\nu\right)}$ has to be estimated. Therefore, the fast orthogonality deficiency compensation, proposed in \cite{Seiler2008}, is applied. The orthogonality deficiency results from the circumstance that even though the basis functions are orthogonal with respect to area $\mathcal{L}$, they are not orthogonal anymore, when evaluated over the support area $\mathcal{A}$ only. Thus, the projection coefficient $p_u^{\left(\nu\right)}$ contains portions from basis functions unlike $\varphi_u\left[m,n\right]$. But, as we want to estimate the portion of $\varphi_u\left[m,n\right]$ in $r^{\left(\nu-1\right)}\left[m,n\right]$ the effect of the orthogonality deficiency has to be compensated. For this purpose $\hat{c}_u^{\left(\nu\right)}$ is estimated as a fraction of $p_u^{\left(\nu\right)}$ according to
\begin{equation}
 	\hat{c}_u^{\left(\nu\right)} = \gamma \cdot p_u^{\left(\nu\right)}. 
\end{equation}
The factor $\gamma$ is from the range between $0$ and $1$ and guarantees that a basis function is not overemphasized in the model. In the case that the portion of the basis function $\varphi_u\left[m,n\right]$ is not covered completely, the same basis function can get selected in a later iteration step again. For a detailed discussion of the orthogonality deficiency problem, please refer to \cite{Seiler2007, Seiler2008}.

To finish an iteration step, the model gets updated according to (\ref{eq:fse_model_update}). The above described iteration steps are repeated, until a predefined number of iterations is reached. Finally, the samples of $g\left[m,n\right]$ corresponding to area $\mathcal{B}$ can be used as estimate for the original signal in this area and can be used to conceal the lost block. 
  

\section{Multiple Selection Extrapolation}
\label{sec:msextrapolation}

In general, the generation of the model based on the correctly received adjacent samples can be regarded as overdetermined problem, since there are more basis functions available than the support area $\mathcal{A}$ has samples. For solving this problem and for estimating the original distribution of basis functions, FSE exploits the sparsity of natural signals. As has been shown in \cite{Candes2007} many natural signals have a sparse representation when regarded with respect to some basis or dictionary. For this reason, the idea of FSE is to determine the strongest present basis function in every iteration step and add this one to the model generated so far. So, inherently a sparse model of the signal is generated.

As has been shown in earlier publications like \cite{Meisinger2004b} and \cite{Seiler2008} this approach is very effective at generating the model. The only shortcoming of FSE is the large number of iterations needed for generating the model and with that its high computational cost. This is due to the fact that in every iteration step only one basis function is selected and added to the model. At this stage, the Multiple Selection Extrapolation (MuSE) comes into play. In many cases, not only one basis function is dominant in the approximation error signal, but several ones, instead. In the case, that several different basis functions might produce similar decrements of the weighted approximation error energy the successive selection used by FSE only is suboptimal. This is due to the fact that the basis functions are not orthogonal when evaluated with respect to area $\mathcal{A}$. By only selecting one basis function in an iteration step, as FSE does, the portions the other basis functions have from the approximation error get modified. For this reason, the same basis functions often have to be selected several times for removing them from the approximation error. To cope with this weakness, the novel Multiple Selection Extrapolation can select several basis functions in one iteration step and the model update is performed for all these basis functions simultaneously. 

Although the idea of iteratively generating a model of the signal is the same, the individual iteration steps differ between MuSE and FSE due to the possibility of multiple selections. Hence the model update has to be modified to
\begin{equation}
	\label{eq:MuSE_model_update}
	g^{\left(\nu\right)}\left[m,n\right] = g^{\left(\nu-1\right)}\left[m,n\right] + \sum_{\forall u\in \mathfrak{G}^{\left(\nu\right)}}\hat{c}^{\left(\nu\right)}_u \cdot \varphi_u\left[m,n\right], 
\end{equation}
where $\mathfrak{G}^{\left(\nu\right)}$ contains the indices of all the basis functions that are added to the model in the $\nu$-th iteration step. The update step of the approximation error $r^{\left(\nu\right)}\left[m,n\right]$ is performed in the same manner:
\[
	r^{\left(\nu\right)}\left[m,n\right] = \hspace{6.9cm}
\]\vspace{-0.55cm}
\begin{equation}
	 \hspace{0.3125cm}\left\{\hspace{-1.75mm} \begin{array}{ll} \displaystyle r^{\left(\nu-1\right)}\left[m,n\right] - \hspace{-4mm} \sum_{\forall u\in \mathfrak{G}^{\left(\nu\right)}}\hspace{-2mm}\hat{c}^{\left(\nu\right)}_u \cdot \varphi_u\left[m,n\right] \hspace{-2mm} & , \forall \left(m,n\right) \in \mathcal{A} \vspace{-0.2cm}\\  0 & , \forall \left(m,n\right) \in \mathcal{B} \end{array} \right.
\end{equation}

To select the basis functions to be added in the $\nu$-th iteration step, again the approximation error $r^{\left(\nu-1\right)}\left[m,n\right]$ is projected onto all basis functions, leading to the projection coefficients as shown in (\ref{eq:projection}). Afterwards, for every basis function, the hypothetical decrement $\Delta \bar{E}_k^{\left(\nu\right)}$ of the weighted approximation error energy, that would be gained if this certain basis function is selected only alone, is computed:
\begin{equation}
	\Delta \bar{E}_k^{\left(\nu\right)} = \gamma^2 \cdot p_k^{\left(\nu\right)^2} \sum_{\left(m,n\right)\in \mathcal{L}} \varphi_k^2\left[m,n\right] \cdot w\left[m,n\right]
\end{equation}\vspace{-2mm}

Therewith, the basis functions that can produce a large decrement of the weighted approximation error energy can be identified. For determining the set $\mathfrak{G}^{\left(\nu\right)}$,
the indices of the basis functions that alone could lead to a decrement larger than the energy fraction threshold $\tau$ times the maximal possible decrement are selected. In addition to that, in order to avoid a too large subspace, the set further is restricted to contain only $N_\mathrm{BF}$ indices at maximum. These are the ones that correspond to the largest $N_\mathrm{BF}$ energy decrements. So, $\mathfrak{G}^{\left(\nu\right)}$ can be written as:\vspace{-0.25cm}
 \[
 	 \hspace{-1cm}\mathfrak{G}^{\left(\nu\right)} = \bigg\{ k \ |\ \Delta \bar{E}_k^{\left(\nu\right)} > \tau \max_{\widetilde{k}} \Delta \bar{E}_{\widetilde{k}}^{\left(\nu\right)} \ \  \wedge
 \]\vspace{-0.35cm}
\begin{equation}
\label{eq:new_bfs}
 \hspace{2.25cm} \Delta \bar{E}_k^{\left(\nu\right)} \geq \mathrm{S_d}\left(\left\{\Delta \bar{E}_{\widetilde{k}}^{\left(\nu\right)}\right\}, N_\mathrm{BF}\right) \bigg\}. 
\end{equation}
Here the function $\mathrm{S_d}\left(\left\{\Delta \bar{E}_{\tilde{k}}^{\left(\nu\right)}\right\}, N_\mathrm{BF}\right)$ returns the \mbox{$N_\mathrm{BF}$-th} element from the decreasing ordered sequence of all the possible approximation error decrements. The restriction of the dimensionality of the subspace is important for keeping the model update efficiently computable. When regarding noise like signals, the number of basis functions satisfying the first condition of (\ref{eq:new_bfs}) becomes very large. Thus, the subsequent computations would get computationally expensive if no upper bound for the number of basis functions is applied.

\begin{figure*}
	\centering
	\vspace{-2.5mm}
	{\psfrag{s01}[b][b]{\color[rgb]{0,0,0}\setlength{\tabcolsep}{0pt}}%
	\psfrag{s02}[t][t]{\color[rgb]{0,0,0}\setlength{\tabcolsep}{0pt}\begin{tabular}{c}Iterations\end{tabular}}%
	\psfrag{s03}[b][b]{\color[rgb]{0,0,0}\setlength{\tabcolsep}{0pt}\begin{tabular}{c}$\PSNR$\end{tabular}}%
	\psfrag{s06}[][]{\color[rgb]{0,0,0}\setlength{\tabcolsep}{0pt}\begin{tabular}{c} \end{tabular}}%
	\psfrag{s07}[][]{\color[rgb]{0,0,0}\setlength{\tabcolsep}{0pt}\begin{tabular}{c} \end{tabular}}%
	\psfrag{s08}[l][l][0.65]{\color[rgb]{0,0,0}\hphantom{``Monarch'':} Goldhill MuSE}%
	\psfrag{s25}[l][l][0.65]{\color[rgb]{0,0,0}``Lena'':\newlength{\nlena}\settowidth{\nlena}{``Lena'':}\hspace{-\nlena}\hphantom{``Monarch'':} FSE}%
	\psfrag{s26}[l][l][0.65]{\color[rgb]{0,0,0}\hphantom{``Monarch'':} MuSE}%
	\psfrag{s27}[l][l][0.65]{\color[rgb]{0,0,0}``Peppers'':\newlength{\npeppers}\settowidth{\npeppers}{``Peppers'':}\hspace{-\npeppers}\hphantom{``Monarch'':} FSE}%
	\psfrag{s28}[l][l][0.65]{\color[rgb]{0,0,0}\hphantom{``Monarch'':} MuSE}%
	\psfrag{s29}[l][l][0.65]{\color[rgb]{0,0,0}``Goldhill'':\newlength{\ngoldhill}\settowidth{\ngoldhill}{``Goldhill'':}\hspace{-\ngoldhill}\hphantom{``Monarch'':} FSE}%
	\psfrag{s30}[l][l][0.65]{\color[rgb]{0,0,0}\hphantom{``Monarch'':} MuSE}%
	\psfrag{x12}[t][t][0.9]{$0$}%
	\psfrag{x13}[t][t][0.9]{$20$}%
	\psfrag{x14}[t][t][0.9]{$40$}%
	\psfrag{x15}[t][t][0.9]{$60$}%
	\psfrag{x16}[t][t][0.9]{$80$}%
	\psfrag{x17}[t][t][0.9]{$100$}%
	\psfrag{x18}[t][t][0.9]{$120$}%
	\psfrag{x19}[t][t][0.9]{$140$}%
	\psfrag{x20}[t][t][0.9]{$160$}%
	\psfrag{x21}[t][t][0.9]{$180$}%
	\psfrag{x22}[t][t][0.9]{$200$}%
	\psfrag{v12}[r][r][0.9]{$22$}%
	\psfrag{v13}[r][r][0.9]{$23$}%
	\psfrag{v14}[r][r][0.9]{$24$}%
	\psfrag{v15}[r][r][0.9]{$25$}%
	\psfrag{v16}[r][r][0.9]{$26$}%
	\psfrag{v17}[r][r][0.9]{$27$}%
	\includegraphics[width=0.4\textwidth]{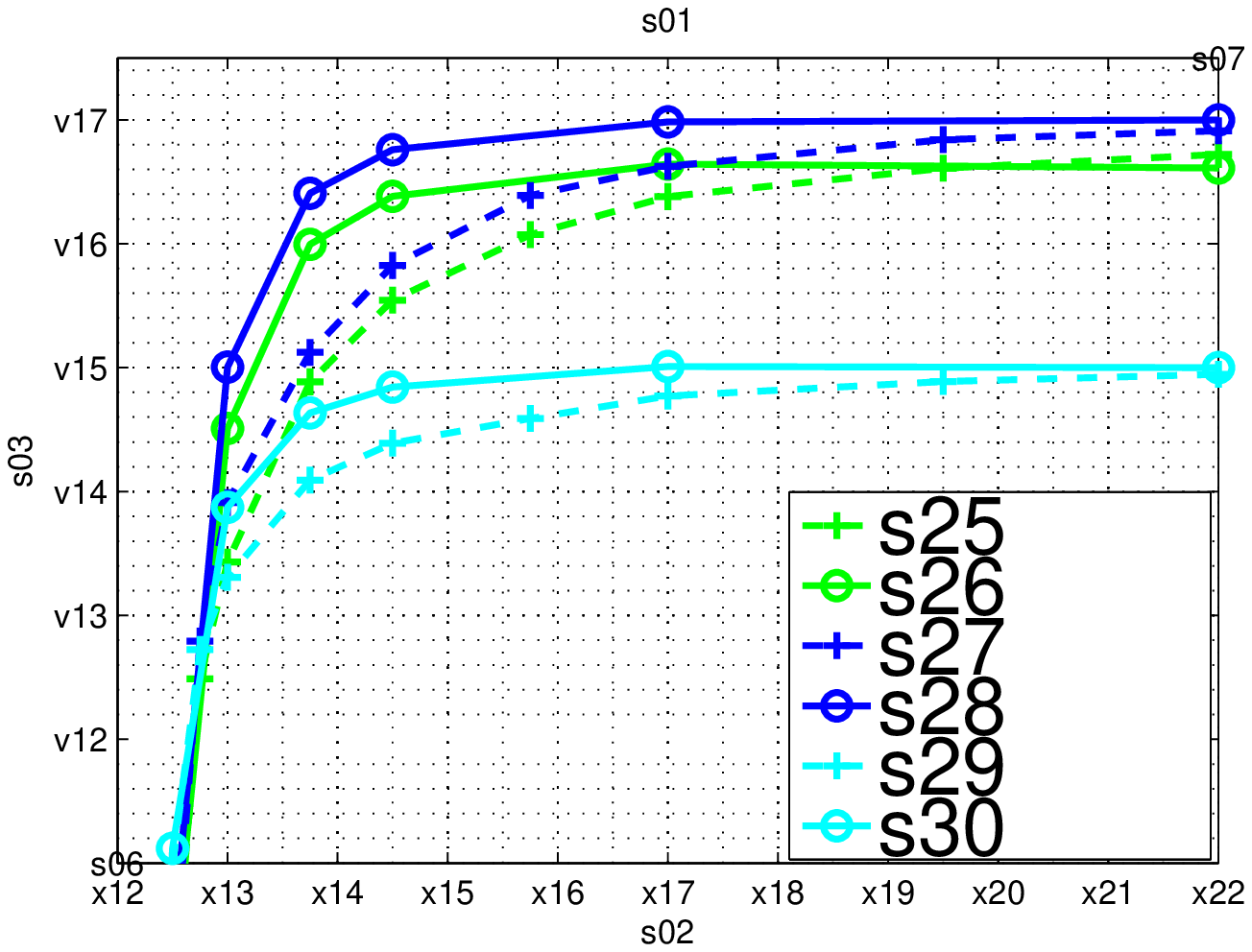}} \hspace{0.75cm}
	{\psfrag{s01}[b][b]{\color[rgb]{0,0,0}\setlength{\tabcolsep}{0pt}}%
	\psfrag{s02}[t][t]{\color[rgb]{0,0,0}\setlength{\tabcolsep}{0pt}\begin{tabular}{c}Iterations\end{tabular}}%
	\psfrag{s03}[b][b]{\color[rgb]{0,0,0}\setlength{\tabcolsep}{0pt}\begin{tabular}{c}$\PSNR$\end{tabular}}%
	\psfrag{s06}[][]{\color[rgb]{0,0,0}\setlength{\tabcolsep}{0pt}\begin{tabular}{c} \end{tabular}}%
	\psfrag{s07}[][]{\color[rgb]{0,0,0}\setlength{\tabcolsep}{0pt}\begin{tabular}{c} \end{tabular}}%
	\psfrag{s08}[l][l][0.65]{\color[rgb]{0,0,0}\hphantom{``Monarch'':} Tulips MuSE}%
	\psfrag{s25}[l][l][0.65]{\color[rgb]{0,0,0}``Baboon'':\newlength{\nbaboon}\settowidth{\nbaboon}{``Baboon'':}\hspace{-\nbaboon}\hphantom{``Monarch'':} FSE}%
	\psfrag{s26}[l][l][0.65]{\color[rgb]{0,0,0}\hphantom{``Monarch'':} MuSE}%
	\psfrag{s27}[l][l][0.65]{\color[rgb]{0,0,0}``Monarch'': FSE}%
	\psfrag{s28}[l][l][0.65]{\color[rgb]{0,0,0}\hphantom{``Monarch'':} MuSE}%
	\psfrag{s29}[l][l][0.65]{\color[rgb]{0,0,0}``Tulips'':\newlength{\ntulips}\settowidth{\ntulips}{``Tulips'':}\hspace{-\ntulips}\hphantom{``Monarch'':} FSE}%
	\psfrag{s30}[l][l][0.65]{\color[rgb]{0,0,0}\hphantom{``Monarch'':} MuSE}%
	\psfrag{x12}[t][t][0.9]{$0$}%
	\psfrag{x13}[t][t][0.9]{$20$}%
	\psfrag{x14}[t][t][0.9]{$40$}%
	\psfrag{x15}[t][t][0.9]{$60$}%
	\psfrag{x16}[t][t][0.9]{$80$}%
	\psfrag{x17}[t][t][0.9]{$100$}%
	\psfrag{x18}[t][t][0.9]{$120$}%
	\psfrag{x19}[t][t][0.9]{$140$}%
	\psfrag{x20}[t][t][0.9]{$160$}%
	\psfrag{x21}[t][t][0.9]{$180$}%
	\psfrag{x22}[t][t][0.9]{$200$}%
	\psfrag{v12}[r][r][0.9]{$17$}%
	\psfrag{v13}[r][r][0.9]{$18$}%
	\psfrag{v14}[r][r][0.9]{$19$}%
	\psfrag{v15}[r][r][0.9]{$20$}%
	\psfrag{v16}[r][r][0.9]{$21$}%
	\psfrag{v17}[r][r][0.9]{$22$}%
	\includegraphics[width=0.4\textwidth]{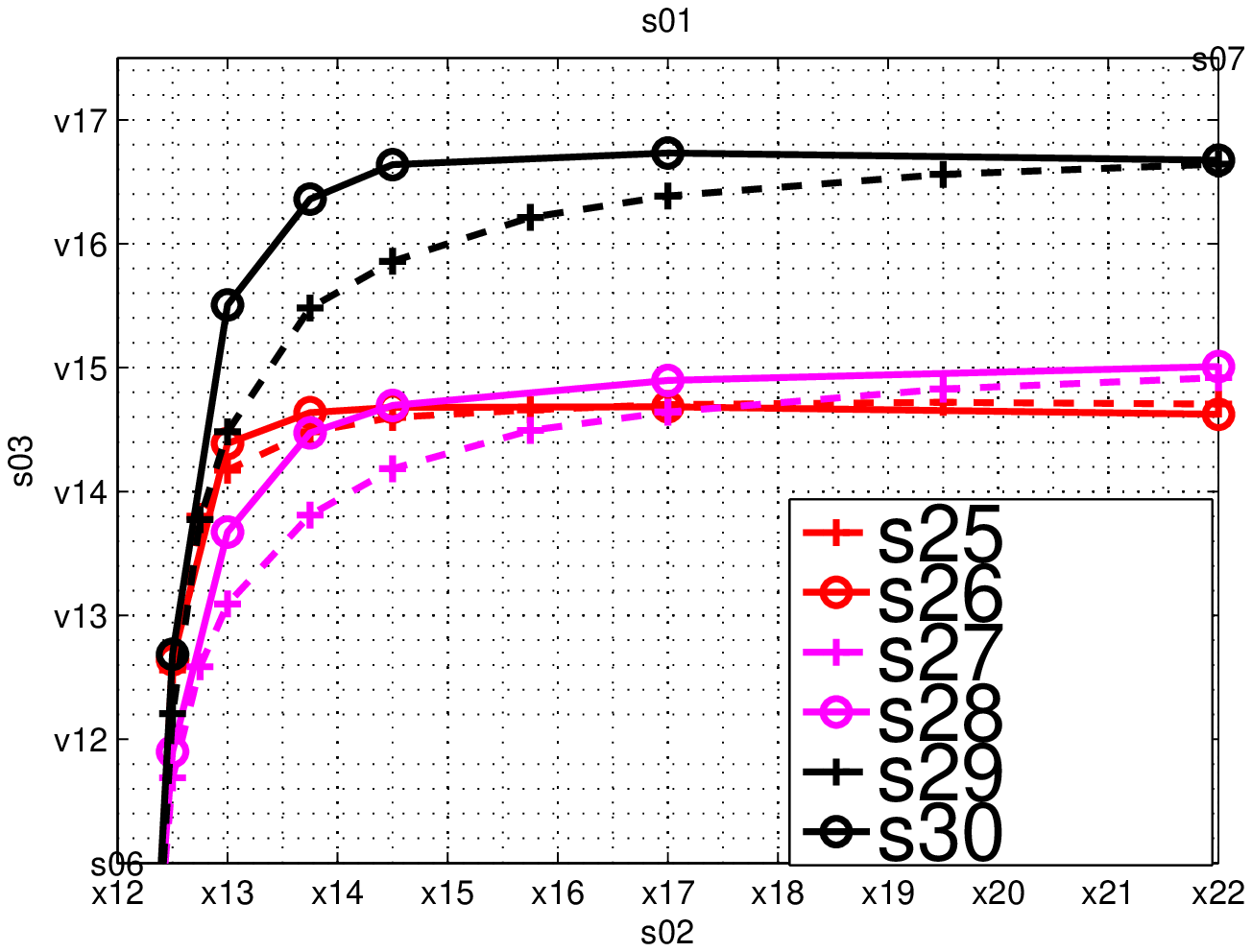}}
	
	\caption{Extrapolation quality over iterations for isolated block losses of size $16\times 16$ samples with $\tau=0.9$ and $N_\mathrm{BF}=5$. Left: test images ``Lena'', ``Peppers'' and `` Goldhill''. Right: test images ``Baboon'', ``Monarch'' and ``Tulips''.\vspace{-4mm}}
	\label{fig:seq_compare}
\end{figure*}

After the set of basis functions has been determined for this iteration step, the approximation error $r^{\left(\nu-1\right)} \left[m,n\right]$ from the previous iteration step is projected onto the subspace spanned by the selected basis functions. For determining the new projection coefficients $\widetilde{p}_u^{\left(\nu\right)}, \forall u \in \mathfrak{G}^{\left(\nu\right)}$ the squared weighted distance 
\begin{equation}
	d^{\left(\nu\right)^2} \hspace{-2mm} =  \hspace{-3mm} \sum_{\left(m,n\right) \in \mathcal{L}} \hspace{-1mm} \left(r^{\left(\nu-1\right)} \left[m,n\right]  \hspace{-1mm}- \hspace{-2mm} \sum_{u\in \mathfrak{G}^{\left(\nu\right)}} \widetilde{p}_u^{\left(\nu\right)} \varphi_u\left[m,n\right] \right)^2  w\left[m,n\right]
\end{equation}
between $r^{\left(\nu-1\right)} \left[m,n\right]$ and the the weighted projection onto the subspace is minimized. This is obtained by setting the partial derivatives of $d^{\left(\nu\right)^2}$ with respect to all $\widetilde{p}_u^{\left(\nu\right)}, \forall u \in \mathfrak{G}^{\left(\nu\right)}$ to zero:
\begin{equation}
	\label{eq:par_div1}
	\frac{\partial d^{\left(\nu\right)^2}}{\partial \widetilde{p}_u^{\left(\nu\right)}} \stackrel{!}{=} 0 \ ,\  \forall u \in \mathfrak{G}^{\left(\nu\right)}
\end{equation}
The partial derivatives with respect to $\widetilde{p}_u^{\left(\nu\right)}$ lead to $\left|\mathfrak{G}^{\left(\nu\right)} \right|$ equations ($\left|\bullet\right|$ depicting the cardinality):
\[
 	\frac{\partial d^{\left(\nu\right)^2}}{\partial \widetilde{p}_u^{\left(\nu\right)}} = \sum_{\left(m,n\right)\in \mathcal{L}} \Bigg( 2 \varphi_u\left[m,n\right] \bigg(r^{\left(\nu-1\right)} \left[m,n\right] - \hspace{1.9cm}
\] \vspace{-0.5cm}
\begin{equation}
	\label{eq:par_div2}
	  \hspace{0.5cm}- \sum_{u\in \mathfrak{G}^{\left(\nu\right)}} \widetilde{p}_u^{\left(\nu\right)} \varphi_u\left[m,n\right] \bigg) w\left[m,n\right] \Bigg)\ , \  \forall u \in \mathfrak{G}^{\left(\nu\right)}.
\end{equation}
For a better differentiation between the summation index and the index of the equations, the equation index $u$ is replaced by $\widetilde{u}$. So, (\ref{eq:par_div1}) and (\ref{eq:par_div2}) can be rewritten and we get the following system of $\left|\mathfrak{G}^{\left(\nu\right)} \right|$ equations that has to be solved for determining the coefficients $\widetilde{p}_u^{\left(\nu\right)}$:
\[
	\sum_{\left(m,n\right)\in \mathcal{L}} \varphi_{\widetilde{u}} \left[m,n\right] r^{\left(\nu-1\right)} \left[m,n\right] w\left[m,n\right] = \hspace{2.75cm}
\]\vspace{-0.25cm}
\begin{equation}
	\label{eq:eqsystem}
	\sum_{u\in \mathfrak{G}^{\left(\nu\right)}} \widetilde{p}_u^{\left(\nu\right)} \sum_{\left(m,n\right)\in \mathcal{L}}  \varphi_{\widetilde{u}} \left[m,n\right]\varphi_u \left[m,n\right] w\left[m,n\right], \ \forall \widetilde{u} \in \mathfrak{G}^{\left(\nu\right)}
\end{equation}
This quadratic linear system of equations can be solved efficiently as the terms $\sum\varphi_{\widetilde{u}}\left[m,n\right]\varphi_u \left[m,n\right] w\left[m,n\right]$ form a symmetric matrix and since the system is maximally of size $N_\mathrm{BF}$ . After all $\widetilde{p}_u^{\left(\nu\right)}$ have been calculated the expansion coefficient update can be derived according to
\begin{equation}
	 \hat{c}_u^{\left(\nu\right)} = \gamma \cdot  \widetilde{p}_u^{\left(\nu\right)} \ , \ \forall u \in \mathfrak{G}^{\left(\nu\right)}.
\end{equation}
The factor $\gamma$ again is used to prevent the model generation for overemphasizing basis functions due to orthogonality deficiency. After this, the model can be updated as shown in (\ref{eq:MuSE_model_update}). Analogous to FSE, these steps are repeated, until a predefined number of iterations is reached. And again, the samples of $g\left[m,n\right]$ corresponding to area $\mathcal{B}$ are an estimate for the original signal that has been lost due to the block loss. 


\section{Simulation Setup and Results}
\label{sec:simulations}

\begin{figure*}
 	\centering
	\includegraphics[width=0.84\textwidth]{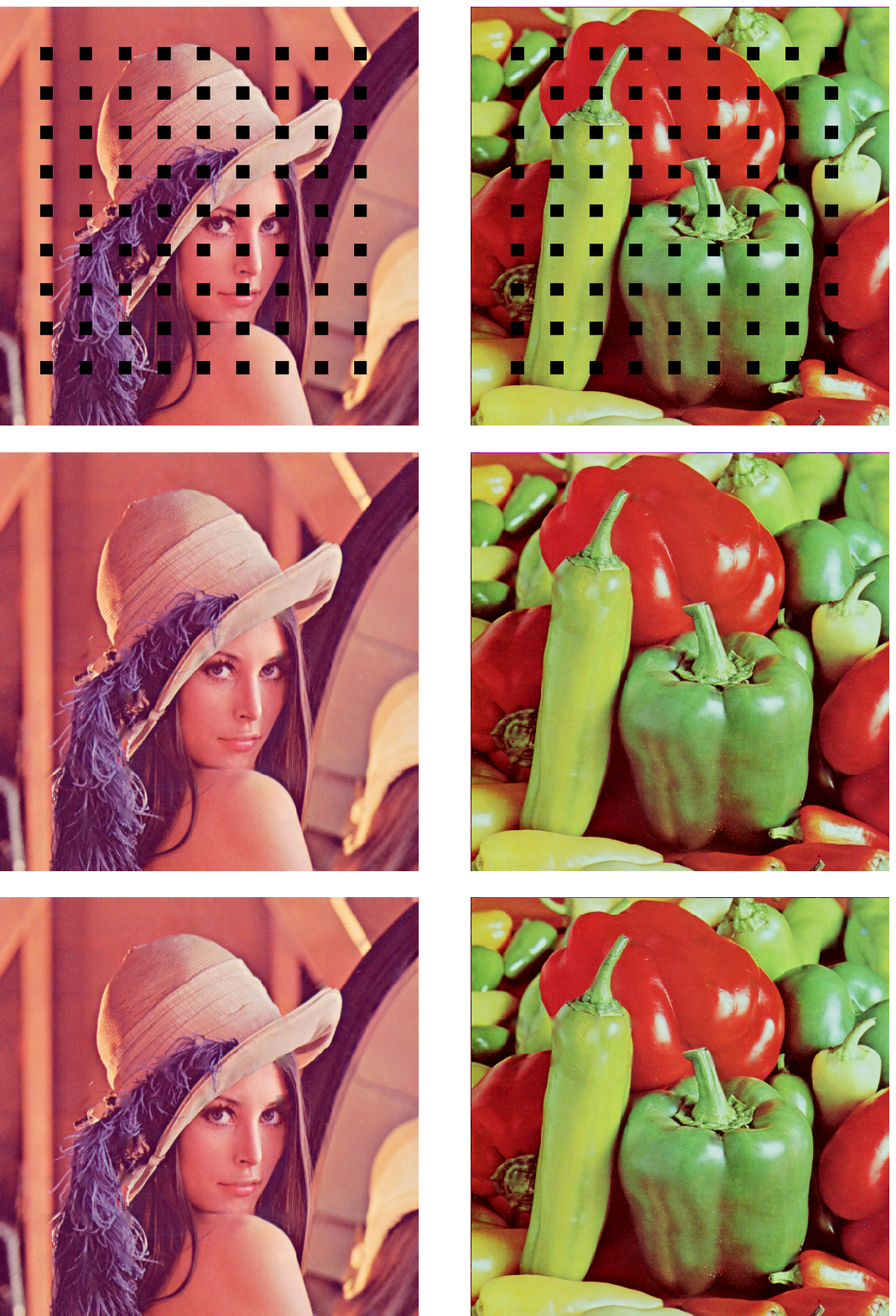}
	\caption{Visual results for test images ``Lena'' (left) and ``Peppers'' (right). Top: error pattern. Mid: error concealment by FSE (\cite{Meisinger2004b, Seiler2008}) with $\gamma=0.2$, $\hat{\rho}=0.8$ and $200$ iterations. Bottom: error concealment by MuSE with $\gamma=0.2$, $\hat{\rho}=0.8, \tau=0.9, N_\mathrm{BF}=5$ and $40$ iterations.}
	\label{fig:visual_results}
\end{figure*}

In order to demonstrate the abilities of Multiple Selection Extrapolation (MuSE) compared to Frequency Selective Extrapolation (FSE) the extrapolation results are evaluated for concealing lost blocks in images. For this, isolated blocks of size $16\times 16$ samples are cut out of the test images ``Baboon'', ``Lena'', ``Peppers'', ``Goldhill'', ``Monarch'' and ``Tulips'' according to the test pattern showed in Fig.\ \ref{fig:visual_results} top. Then, these blocks are extrapolated by MuSE and FSE and are compared to the original blocks by calculating the $\PSNR$. Without loss of generality, only isolated block losses are regarded, since the concealment procedure can easily be extended to more complex loss patterns as well, as described in \cite{Meisinger2004b}. Thereby, larger losses are divided into small blocks which are concealed individually. In doing so, the already concealed blocks are included into the support area of the unconcealed ones, but get a lower weight in the weighting function for reducing error propagation.

The extrapolation parameters, that are valid for both, MuSE and FSE, are chosen according to \cite{Seiler2008}. So, the support area is a frame of $16$ samples around the the lost block. The weighting function 
\begin{equation}
 	\rho\left[m,n\right] = \hat{\rho}^{\sqrt{\left(m-\frac{M-1}{2}\right)^2 + \left(n-\frac{N-1}{2}\right)^2}}
\end{equation}
emanates from a radial symmetric isotropic model that causes an exponentially decreasing weight with an increasing distance from the center. Hence, samples far away from the lost block only get a low weight and therewith only small influence on the model generation. According to \cite{Seiler2008}, the decay factor $\hat{\rho}$ is chosen to $0.8$ and the orthogonality deficiency compensation factor $\gamma$ is set to $0.2$. The set of used basis functions is the set of the two-dimensional discrete Fourier transform. As has been pointed out in earlier publications as e.\ g. \cite{Seiler2008, Meisinger2004b, Meisinger2007} this set of basis functions is especially suited for concealment of losses in natural images since monotone as well as noise like areas and edges can be reconstructed well. In addition to that, by using this set of basis functions many of the above outlined calculations can be performed efficiently in the transform domain. For a detailed discussion of the frequency domain implementation, please refer to \cite{Meisinger2004b}. Other sets of basis functions as e.\ g.\ the one emanating from the Discrete Cosine Transform can be used as well, but have shown a slightly lower performance for error concealment compared to the Fourier basis functions.

Since the projection on a complete subspace should only be applied in the case that several basis functions could lead to similar decrements of the weighted approximation error energy, the energy fraction threshold $\tau$ is chosen to $0.9$. In order to keep the system of equations from (\ref{eq:eqsystem}) manageable and to prevent the algorithm from selecting too many basis functions in one iteration step, the maximum number $N_\mathrm{BF}$ of basis functions to be added in an iteration step is set to $5$. 

Fig.\ \ref{fig:seq_compare} shows the extrapolation results for the above mentioned test images with respect to the number of iterations by using FSE and MuSE. Comparing the two graphs belonging to one image, one can easily see that MuSE has a significantly faster ascent of $\PSNR$ compared to FSE. So, fewer iterations are needed to obtain a certain extrapolation quality. For small numbers of iterations the gain of MuSE over FSE can be more than $1 \punit{dB}$ $\PSNR$. For a large number of iterations the graphs of MuSE and FSE converge and run into saturation reaching approximately the same extrapolation quality. To quantify the speed-up obtainable by MuSE, Table \ref{tab:iterations} shows the number of iterations needed for reaching saturation quality by $0.25 \punit{dB}$ for MuSE and FSE. The offset of $0.25 \punit{dB}$ is used since the actual number of iterations needed for obtaining saturation cannot be determined accurately as the curves become horizontal there. Comparing the listed values, it becomes apparent, that the number of iterations can be reduced by a factor of up to more than $3$ for the chosen parameters. For small values of $N_\mathrm{BF}$, this factor can be approximately adopted for the reduction in processing time as well, since the modifications of MuSE compared to FSE are computationally not very expensive. During one iteration, the most expensive step is the approximation error's projection onto all the $M$ times $N$ basis functions according to (\ref{eq:projection}). There, for every projection the weighted scalar product between the approximation error and the basis function has to be evaluated. Compared to this, the additional complexity of MuSE added by determining the set $\mathfrak{G}^{\left(\nu\right)}$ and by solving the small system of equations from (\ref{eq:eqsystem}) is negligible. So, as the computationally most expensive step has to be carried out for FSE as well as MuSE, the overall computational cost per iteration is similar for both algorithms.

\begin{table}
 	\centering
	\caption{Iterations needed for obtaining saturation quality by $0.25 \punit{dB}$}
	\label{tab:iterations}
	\begin{tabular}{|c|c|c|c|}
	 	\hline & FSE \cite{Meisinger2004b,Seiler2008} & MuSE & Ratio\\
		\hline ``Baboon'' & $32$ & $20$ & $ 1.6$ \\
		\hline ``Lena'' & $104$ & $51$ & $ 2.0$ \\
		\hline ``Peppers'' & $130$ & $50$ & $ 2.6$ \\
		\hline ``Goldhill'' & $98$ & $45$ & $ 2.2$ \\
		\hline ``Monarch'' & $132$ & $65$ & $ 2.0$ \\
		\hline ``Tulips'' & $129$ & $42$ & $ 3.1$ \\ \hline 
	\end{tabular}\vspace{-2.5mm}
\end{table}

\begin{figure}
	\centering
	\psfrag{s01}[b][b]{\color[rgb]{0,0,0}\setlength{\tabcolsep}{0pt}}%
	\psfrag{s02}[t][t]{\color[rgb]{0,0,0}\setlength{\tabcolsep}{0pt}\begin{tabular}{c}Iterations\end{tabular}}%
	\psfrag{s03}[b][b]{\color[rgb]{0,0,0}\setlength{\tabcolsep}{0pt}\begin{tabular}{c}$\PSNR$\end{tabular}}%
	\psfrag{s06}[][]{\color[rgb]{0,0,0}\setlength{\tabcolsep}{0pt}\begin{tabular}{c} \end{tabular}}%
	\psfrag{s07}[][]{\color[rgb]{0,0,0}\setlength{\tabcolsep}{0pt}\begin{tabular}{c} \end{tabular}}%
	\psfrag{s08}[l][l][0.65]{\color[rgb]{0,0,0}``Goldhill:'' $\tau=0.95$, $N_\mathrm{BF}=5$}%
	\psfrag{s13}[l][l][0.65]{\color[rgb]{0,0,0}``Peppers:''\newlength{\nnpeppers}\settowidth{\nnpeppers}{``Peppers:'':}\hspace{-\nnpeppers}\hphantom{``Peppers:'' } $\tau=0.90$, $N_\mathrm{BF}=5$}%
	\psfrag{s14}[l][l][0.65]{\color[rgb]{0,0,0}\hphantom{``Goldhill'':} $\tau=0.90$, $N_\mathrm{BF}=3$}%
	\psfrag{s15}[l][l][0.65]{\color[rgb]{0,0,0}\hphantom{``Goldhill'':} $\tau=0.90$, $N_\mathrm{BF}=7$}%
	\psfrag{s16}[l][l][0.65]{\color[rgb]{0,0,0}\hphantom{``Goldhill'':} $\tau=0.75$, $N_\mathrm{BF}=5$}%
	\psfrag{s17}[l][l][0.65]{\color[rgb]{0,0,0}\hphantom{``Goldhill'':} $\tau=0.95$, $N_\mathrm{BF}=5$}%
	\psfrag{s18}[l][l][0.65]{\color[rgb]{0,0,0}``Goldhill'': $\tau=0.90$, $N_\mathrm{BF}=5$}%
	\psfrag{s19}[l][l][0.65]{\color[rgb]{0,0,0}\hphantom{``Goldhill'':} $\tau=0.90$, $N_\mathrm{BF}=3$}%
	\psfrag{s20}[l][l][0.65]{\color[rgb]{0,0,0}\hphantom{``Goldhill'':} $\tau=0.90$, $N_\mathrm{BF}=7$}%
	\psfrag{s21}[l][l][0.65]{\color[rgb]{0,0,0}\hphantom{``Goldhill'':} $\tau=0.75$, $N_\mathrm{BF}=5$}%
	\psfrag{s22}[l][l][0.65]{\color[rgb]{0,0,0}\hphantom{``Goldhill'':} $\tau=0.95$, $N_\mathrm{BF}=5$}%
	\psfrag{x12}[t][t][0.9]{$0$}%
	\psfrag{x13}[t][t][0.9]{$20$}%
	\psfrag{x14}[t][t][0.9]{$40$}%
	\psfrag{x15}[t][t][0.9]{$60$}%
	\psfrag{x16}[t][t][0.9]{$80$}%
	\psfrag{x17}[t][t][0.9]{$100$}%
	\psfrag{x18}[t][t][0.9]{$120$}%
	\psfrag{x19}[t][t][0.9]{$140$}%
	\psfrag{x20}[t][t][0.9]{$160$}%
	\psfrag{x21}[t][t][0.9]{$180$}%
	\psfrag{x22}[t][t][0.9]{$200$}%
	\psfrag{v12}[r][r][0.9]{$22$}%
	\psfrag{v13}[r][r][0.9]{$23$}%
	\psfrag{v14}[r][r][0.9]{$24$}%
	\psfrag{v15}[r][r][0.9]{$25$}%
	\psfrag{v16}[r][r][0.9]{$26$}%
	\psfrag{v17}[r][r][0.9]{$27$}%
	\psfrag{v18}[r][r]{}%
	\vspace{-5mm}
	\includegraphics[width=0.4\textwidth]{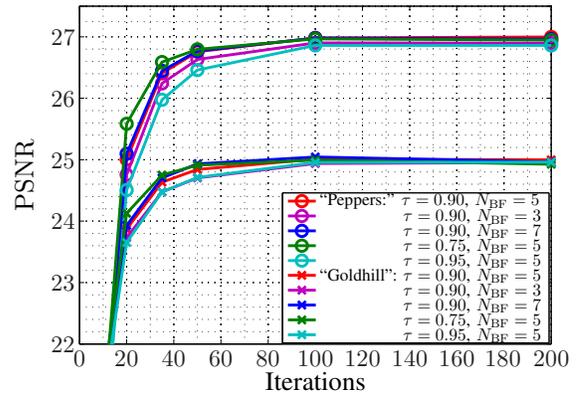}
	\caption{Extrapolation quality over iterations for isolated block losses of size $16\times 16$ samples with different values for $\tau$ and $N_\mathrm{BF}$.\vspace{-4mm}}
	\label{fig:parameter_compare}
\end{figure}

The above chosen values of $\tau$ and $N_\mathrm{BF}$ fortunately are uncritical and can be varied in a relatively wide range without affecting the extrapolation quality. To prove this, Fig.\ \ref{fig:parameter_compare} shows the $\PSNR$ over iterations with different combinations of $\tau$ and $N_\mathrm{BF}$ for the test images ``Peppers'' and ``Goldhill''. Although the graphs are widened a little bit for small numbers of iterations, they all show the same behavior and all lead to a comparable saturation quality. By adjusting $\tau$ and $N_\mathrm{BF}$ one can also easily trade extrapolation quality against extrapolation speed and tune the algorithm for the desired application.

Besides the objective extrapolation results, the visual extrapolation quality is very important for error concealment. To demonstrate this, Fig.\ \ref{fig:visual_results} shows the results for concealment of isolated block losses for the test images ``Lena'' and ``Peppers''. There, the concealment with FSE was carried out with $200$ iterations and for MuSE $40$ iterations were applied. Comparing the mid and bottom row of the images, it becomes apparent, that already a fifth of the iterations is needed to obtain the same high visual extrapolation quality with MuSE compared to FSE.

\section{Conclusion}
\label{sec:conclusion}

In the scope of this paper a novel extrapolation algorithm, the Multiple Selection Extrapolation, was presented. This algorithm is an improvement to the already efficient Frequency Selective Extrapolation. Compared to this algorithm the Multiple Selection Extrapolation significantly reduces the overall computational cost and the therewith the processing time for extrapolation. At the same time, the very high subjective as well as objective extrapolation quality is preserved. In addition to this, the novel algorithm can effectively trade extrapolation quality against extrapolation speed, in order to find the best compromise for a desired application. Furthermore, although MuSE was outlined for spatial error concealment only, due to its relationship to FSE the algorithm can be extended to three- and four-dimensional data sets as well by using the concepts presented in \cite{Meisinger2007, Fecker2008}. Therewith, it can easily be applied to error concealment in video and multiview sequences, making full use of the there available data.


\clearpage


\begin{thebibliography}{10}
	\providecommand{\url}[1]{#1}
	\def\UrlFont{\rmfamily}
	\providecommand{\newblock}{\relax}
	\providecommand{\bibinfo}[2]{#2}
	\providecommand\BIBentrySTDinterwordspacing{\spaceskip=0pt\relax}
	\providecommand\BIBentryALTinterwordstretchfactor{4}
	\providecommand\BIBentryALTinterwordspacing{\spaceskip=\fontdimen2\font plus
		\BIBentryALTinterwordstretchfactor\fontdimen3\font minus
		\fontdimen4\font\relax}
	\providecommand\BIBforeignlanguage[2]{{%
			\expandafter\ifx\csname l@#1\endcsname\relax
			\typeout{** WARNING: IEEEtran.bst: No hyphenation pattern has been}%
			\typeout{** loaded for the language `#1'. Using the pattern for}%
			\typeout{** the default language instead.}%
			\else
			\language=\csname l@#1\endcsname
			\fi
			#2}}
	
	\bibitem{Wang1993}
	Y.~Wang, Q.-F. Zhu, and L.~Shaw, ``Maximally smooth image recovery in transform
	coding,'' \emph{IEEE Trans. on Commun.}, vol.~41, no.~10, pp. 1544--1551,
	Oct. 1993.
	
	\bibitem{Alkachouh2000}
	Z.~Alkachouh and M.~Bellanger, ``Fast {DCT}-based spatial domain interpolation
	of blocks in images,'' \emph{IEEE Trans. Image Process.}, vol.~9, no.~4, pp.
	729--732, April 2000.
	
	\bibitem{Sun1995}
	H.~Sun and W.~Kwok, ``Concealment of damaged block transform coded images using
	projections onto convex sets,'' \emph{IEEE Trans. Image Process.}, vol.~4,
	no.~4, pp. 470--477, April 1995.
	
	\bibitem{Li2002}
	X.~Li and M.~T. Orchard, ``Novel sequential error-concealment techniques using
	orientation adaptive interpolation,'' \emph{IEEE Trans. Circuits Syst. Video
		Technol.}, vol.~12, no.~10, pp. 857--864, Oct. 2002.
	
	\bibitem{Zhao2008}
	Y.~Zhao, Q.~Chen, H.~Chen, and M.~Rupp, ``Spatial error concealment using
	optimized directional decision and extrapolation,'' in \emph{Visual
		Information Engineering, 2008. VIE 2008. 5th International Conference on},
	29.07 - 01.08 2008, pp. 658--661.
	
	\bibitem{Meisinger2004b}
	K.~Meisinger and A.~Kaup, ``Minimizing a weighted error criterion for spatial
	error concealment of missing image data,'' in \emph{Proc. Int. Conf. on Image
		Processing (ICIP)}, Singapore, 24.-27. Oct. 2004, pp. 813--816.
	
	\bibitem{Seiler2008}
	J.~Seiler and A.~Kaup, ``Fast orthogonality deficiency compensation for
	improved frequency selective image extrapolation,'' in \emph{Proc. Int. Conf.
		on Acoustics, Speech, and Signal Processing (ICASSP)}, Las Vegas, USA, 31.
	March - 4. April 2008, pp. 781--784.
	
	\bibitem{Seiler2007}
	J.~Seiler, K.~Meisinger, and A.~Kaup, ``Orthogonality deficiency compensation
	for improved frequency selective image extrapolation,'' in \emph{Proc.
		Picture Coding Symposium (PCS)}, Lisboa, Portugal, 7.-9. Nov. 2007.
	
	\bibitem{Candes2007}
	E.~Cand\`{e}s, N.~Braun, and M.~Wakin, ``Sparse signal and image recovery from
	compressive samples,'' in \emph{Biomedical Imaging: From Nano to Macro, 2007.
		ISBI 2007. 4th IEEE International Symposium on}, April 2007, pp. 976--979.
	
	\bibitem{Meisinger2007}
	K.~Meisinger and A.~Kaup, ``Spatiotemporal selective extrapolation for 3-{D}
	signals and its applications in video communications,'' \emph{IEEE Trans.
		Image Process.}, vol.~16, no.~9, pp. 2348--2360, Sept. 2007.
	
	\bibitem{Fecker2008}
	U.~Fecker, J.~Seiler, and A.~Kaup, ``4-{D} frequency selective extrapolation
	for error concealment in multi-view video,'' in \emph{Proc. International
		Workshop on Multimedia Signal Processing (MMSP)}, Cairns, Australia, 8.-10.
	Oct. 2008, pp. 267--272.
	
\end{thebibliography}
\end{document}